\documentclass[twocolumn,showpacs,preprintnumbers,amsmath,amssymb]{revtex4}
\usepackage{graphicx}% Include figure files
\usepackage{dcolumn}% Align table columns on decimal point
\usepackage{bm}% bold math

\newcommand{\sigbfpp}{\sigma_{\mathrm{\small{BFPP}}}}

\newcommand{\pbtwo}{^{208}\mathrm{Pb}^{82+}}

\begin{document}

%\preprint{APS/123-QED}

\title{Observations of beam losses due to\\bound-free pair production in a heavy-ion collider}

\author{R.~Bruce}
 \altaffiliation[Also at ]{MAXlab, Lund University, Sweden.}%Lines break automatically or can be forced with \\
 \email{roderik.bruce@cern.ch}
\author{J.M.~Jowett}
\author{S.~Gilardoni}
\affiliation{CERN, Geneva, Switzerland}

\author{A.~Drees}
\author{W.~Fischer}
\author{S.~Tepikian}
\affiliation{BNL, Upton, NY, USA}%

\author{S.R. Klein}
\affiliation{LBNL, Berkeley, CA, USA}

\date{\today}% It is always \today, today,

\begin{abstract}

We report the first observations of beam losses due to bound-free pair
production at the interaction point of a heavy-ion collider.  This
process is expected to be a major luminosity limit for the Large Hadron Collider (LHC) when it
operates with $\pbtwo$ ions because the localized energy deposition by
the lost ions may quench superconducting magnet coils.  Measurements
were performed at the BNL Relativistic Heavy Ion Collider (RHIC) during operation
with 100~GeV/nucleon
$^{63}$Cu$^{29+}$ ions. At RHIC, the
rate, energy and magnetic field are low enough so that magnet quenching is not an issue.
The hadronic showers produced
when the single-electron ions struck the RHIC beampipe were observed
using an array of photodiodes.
The measurement confirms the order of magnitude of the
theoretical cross section previously calculated by others.

\end{abstract}

\maketitle

When fully-stripped heavy ions of atomic numbers $Z_1$, $Z_2$ are
brought into collision at the interaction point (IP) of a collider, a
number of electromagnetic interactions are induced by the intense fields
generated by the coherent action of all the $Z_{1,2}$ charges in either
nucleus (for a review, see Ref.~\cite{bert05}).  Some of these
``ultra-peripheral'' interactions have much higher cross sections than
the hadronic nuclear interactions that are the main object of study.
Among them, the bound-free pair production (BFPP), sometimes known as
electron capture from pair production, occurs when the virtual
photon exchanged by the ions converts into a pair, and the electron is
created in an atomic shell of one of the ions:
\begin{equation}
Z_1  + Z_2
\stackrel{\gamma}{\longrightarrow}
(Z_1  + \mathrm{e}^-)_{1s_{1/2},...}  + Z_2  + \mathrm{e}^+
\end{equation}
The resulting one-electron atoms have a slightly larger magnetic
rigidity than the original bare nucleus.
(Magnetic rigidity is defined as $p/(Qe)=B\rho$
for a particle with momentum $p$ and charge $Qe$ that
would have bending radius $\rho$ in a magnetic field $B$).  Since the
transverse recoil is very small, this ``secondary beam'' will emerge at
a very small angle to the main beam.  However it will be bent and
focused less by the guiding magnetic elements and may be lost
somewhere in the collider ring.

It has long been known that this process, together with electromagnetic
dissociation of the nuclei, could be a major contribution to the
intensity and luminosity decay of ion colliders~\cite{gould84,baltz96}.
It was realized more recently~\cite{klein01,pac2003} that, in certain
conditions, the BFPP beam will be lost in a well-defined spot,
initiating hadronic showers in the vacuum envelope of the beam.
The resulting localized heat deposition could induce quenches of
superconducting magnets.  Detailed calculations have been given elsewhere
for 2.76~TeV/nucleon $\pbtwo$ operation of the LHC at
CERN~\cite{lhcdesign,pac2003,pac2005,note379}. The
consequent luminosity limit is expected to occur at a level close to
the design performance.
It is therefore vital to test our quantitative understanding of the
features of the BFPP process in order to ensure safe operation of the
LHC, uninterrupted by lengthy quench-recovery procedures.  BFPP
has been measured in fixed target
experiments~\cite{belkacem93,krause98,grafs99} at lower energy but
not, until now, in a collider.  An opportunity arose during
$^{63}$Cu$^{29+}$ operation of RHIC at Brookhaven National Laboratory.

The flux of ions in the secondary BFPP beam emerging from the interaction point is the
product of the collider luminosity and the BFPP cross section. The partial
cross section for electron capture for a given low-lying atomic bound state~$i$ on
 nucleus~1 has the approximate form~\cite{meier01}
\begin{equation}
\label{eq:sigmabfpp}
\sigma_i \approx Z_1^5  Z_2^2  \left(A_i  \log\gamma_{\mathrm{cm}} + B_i\right)
\end{equation}
where $\gamma_{\mathrm{cm}}$ is the Lorentz boost of the ions in the
centre-of-mass frame and $A_i,B_i$ depend only weakly on $Z$.
The total cross section for BFPP to
\emph{one} of the colliding ions is then
\begin{equation}
\sigbfpp=\sum_i{\sigma_i}
\end{equation}
where the sum is over all atomic shells.
Some numerical values from~\cite{meier01} are given in Table~\ref{tab:BFPP}.
The cross section for collisions of 100~GeV/nucleon $^{63}$Cu$^{29+}$  was not
calculated in Ref.~\cite{meier01} but we have estimated it as 0.2~barn by
interpolation of the data given in
Fig.~7 of Ref.~\cite{meier01}, scaled with $Z$ and $\log \gamma_{\mathrm{cm}}$
according to Eq.~(\ref{eq:sigmabfpp}).
Applied to other tabulated values in Ref.~\cite{meier01},
this method produces an agreement within 10\%.

\begin{table}
\caption{\label{tab:BFPP}BFPP cross sections, typical peak luminosity,
BFPP rates and relative changes in magnetic rigidity at RHIC and LHC.
Values are taken directly where possible, or estimated by fitting sums
of contributions of the form (\ref{eq:sigmabfpp}), to the information
in Ref.~\cite{meier01}.  $\delta=1/(Z-1)$ is the fractional deviation of the
magnetic rigidity.}
\begin{ruledtabular}
\begin{tabular}{|c|c|c|c|c|}
& $\sigbfpp$& $L/10^{27} $ &BFPP & $\delta (\%)$\\
& (barn) & (cm$^{-2}$s$^{-1})$ & rate (kHz) & \\
\hline
LHC Pb-Pb & 281 & 1 & 281 & 1.2\\
2759 GeV/nucleon & & & & \\
\hline
RHIC Au-Au & 114 & 3 & 342 & 1.3 \\
100 GeV/nucleon  & & & &  \\
\hline
RHIC Cu-Cu & 0.2 & 20 & 4 & 3.6\\
100 GeV/nucleon  & & & &  \\
\hline
RHIC Cu-Cu & 0.08 & 1 & 0.08 & 3.6\\
31 GeV/nucleon  & & & &  \\
\end{tabular}
\end{ruledtabular}
\end{table}

The impact point of the lost ions is predicted by tracking the orbit of
the BFPP beam in the collider optics until it intersects the physical
aperture (vacuum pipe) of the beam line.
To lowest order in the uncorrelated betatron amplitudes,
$a$, $b$, and
relative magnetic rigidity deviation from the central value, $\delta$,
the horizontal displacement from the
central orbit, at a distance $s$ from the IP is
\begin{equation}
\label{eq:HillSol}
x(s)= a \sqrt{2 \beta(s)}\cos\left[\mu (s)\right]
      +b \sqrt{2 \beta(s)}\sin\left[\mu(s)\right]+d_x(s)\delta.
\end{equation}
Averaging over all ions gives the transverse emittance
of the beam: $\langle a^2+b^2 \rangle=\epsilon$;
$\beta(s)$ is the usual ``Twiss
function'' determined by the focusing properties of the accelerator,
$\mu(s)=\int_0^s \beta(u)^{-1}d u$
is the betatron phase and $d_x(s)$ is
the locally generated dispersion
($d_x(0)=d^\prime_x(0)=0$)
that encodes the spectrometer effect of the bending magnets.

An ion following the central trajectory (i.e. in the middle of the bunch) in the BFPP beam enters the IP with
$a=b=\delta=0$ and emerges, still with
$a=b=0$, but $\delta=1/(Z-1)$.
This ion will hit the beam pipe at the first location where the horizontal
aperture $A_x$ satisfies $A_x(s)=d_x(s)/(Z-1) $. In the arc, $A_x=3.455$~cm.

\begin{figure}
  \includegraphics[width=8.5cm]{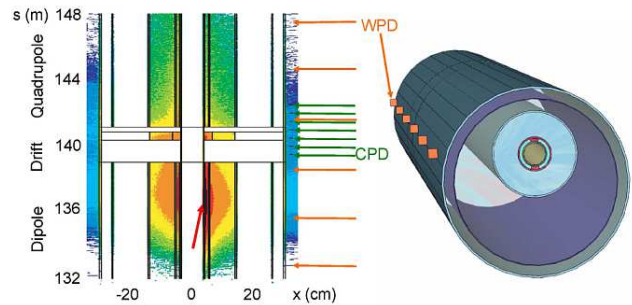}\\
  \caption{(color online) Left: The energy deposition from a typical shower shown in a
thin slice in the $x-s-$plane through the geometry. The red arrow indicates the impact point and
the green and orange arrows show the positions of the PDs in the CPD and WPD. Right: The 3D model
of the geometry as implemented in FLUKA around the impact point. The orange squares show the PDs in the WPD.}\label{fig:shower}
\end{figure}

During $^{197}$Au$^{79+}$ operation at RHIC, $\delta$ is not large enough
for this to occur ($x_\mathrm{max}\simeq 2.5$~cm).  However, during
$^{63}$Cu$^{29+}$ runs~\cite{pilat05}, the larger $\delta$
(Table~\ref{tab:BFPP}) meant that the ions should be lost in a well-defined
location where they can be detected directly.
Thanks to the small cross section, there was no risk of quenching
any superconducting materials.
Since the BFPP beam was not visible to the existing
detectors,
beam loss monitors in the form of PIN
diodes (PDs) of the type Hamamatsu~S3590 were mounted on the outside of the magnet
cryostat downstream of the IP of the PHENIX experiment. The PDs have a time resolution of 25~ns and the counts were
read out every second. They were initially positioned in a wide configuration
(WPD), 3~m apart (see Fig.~\ref{fig:shower}), and later moved closer together (CPD), 0.5~m apart,
around an observed count-rate maximum at 141.6~m.

\begin{figure}
\includegraphics[width=8.5cm]{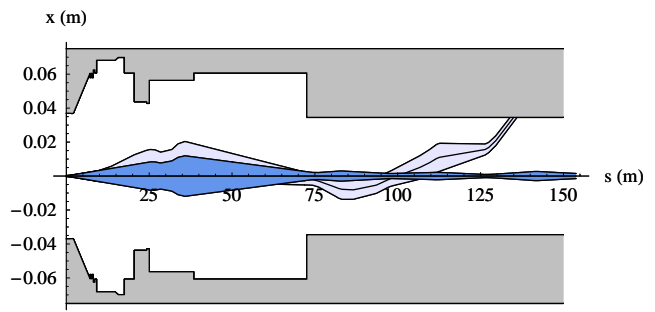}
\caption{\label{fig:Envelope} (color online) The horizontal projection of the 1 $\sigma$ envelope of the BFPP beam
and the nominal beam emerging from the PHENIX IP. The BFPP beam hits the inside of the vacuum
chamber at 135.5~m.}
\end{figure}

A linear interpolation of $d_x(s)$ between magnet ends, based on the
design model of the collider, identified the
elements within which the BFPP beam was lost: a 9.7~m long dipole magnet
starting at    $s=129.6\,\mathrm{m}$,
followed by a 2.12~m drift space and a quadrupole magnet, shown in Fig~\ref{fig:shower}. More precise
tracking, including chromatic effects ($\delta$-dependences), and a
final step of analytic orbit calculation inside the dipole, predicted an
impact at $s_i\simeq135.5$~m
with an angle of incidence of
$\langle x^\prime(s_i)\rangle \simeq\mbox{2.7 mrad}$.
Tracking an assumed Gaussian distribution of
the amplitudes  $a$ and $b$ in Eq.~(\ref{eq:HillSol}) gives the
distribution of impact momenta in the spot around this point.
Fig.~\ref{fig:Envelope} shows the main
$^{63}$Cu$^{29+}$ beam together with the
$^{63}$Cu$^{28+}$ BFPP beam propagating from the IP.

To the extent that the RHIC hardware differs from the design model,
the impact point may be shifted.
The global distortion of the central orbit  by
misalignments of the magnets is partially corrected by
localized corrector magnets.
A least-squares fit was made to orbit data from beam position monitors (BPMs) and
measured misalignments (of order 1--2~mm) of the quadrupole
magnets in the form
\begin{eqnarray}
\label{eq:xcorr}
x(s)&=&x_\beta(s)+\sum_j{\Theta\left(s-s_j\right)\times}\nonumber\\
              &\times &\sqrt{\beta(s)\beta(s_j)}
                        \sin\left[\mu(s)-\mu(s_j)\right]\theta_j.
\end{eqnarray}
Here $\Theta(s)$ is the unit step function, $x_\beta$ the first two
terms in Eq.~(\ref{eq:HillSol}) and $\theta_j$ are the angular kicks
from misalignments and correctors at locations $s_j$.
Unfortunately not enough data were available to make a satisfactory fit.
Instead, the possible spread of the impact point was estimated by calculating
the propagated error  using Eq.~(\ref{eq:xcorr}).
A possible error on the
horizontal quadrupole misalignments of 1~mm was assumed, corresponding to the 95~\%
confidence interval of the measurement and the fact that
the magnets may have moved between early 2005 when the BFPP measurement
was done and late 2006 when the misalignments were measured.
The absolute error on the BPM data  was taken to be 0.5~mm, reflecting
the limited data and possible non-reproducibility.
Combining these  errors gives an impact point of $135.5\pm2$~m. Keeping the nominal optics, but
displacing the beam pipe by 1.4~mm at the impact point according to the measured misalignment of the
dipole, moves the impact point to 136.0~m.

Several distributions of BFPP ions at the PHENIX IP, centered on different initial offsets and angles,
were tracked to obtain their impact coordinates on the inside of the vacuum pipe. These were fed into a
3D model (see Fig.~\ref{fig:shower}) of the impact region geometry,  the PDs and the dipole magnetic field, implemented in the FLUKA~2006.3 Monte Carlo code~\cite{fluka1,fluka2,roes00}, in order to simulate the shower in the magnet and particles emerging from the cryostat. This code has been benchmarked by others in the relevant energy range~\cite{roes00}.

Within 2--2.5~m from the average impact point most of the shower inside the magnet has died out, although
particles outside propagate farther. This can be understood through a rough estimate: The typical average
nuclear interaction length for the superconducting coil and the iron cold bore is about 20~cm.
Convoluted with the impact point spot size of about 1.8 m, this gives a total shower length of about 2.8~m,
assuming that the shower lasts five interaction lengths. The maximum PD signal was expected about 2.5~m
downstream of the impact point. When the  shower is contained within the dipole, moving the impact point
along $s$ only translates the shower. If the impact point is moved closer towards the end of the magnet
or the angle of incidence decreased, more of the shower emerges into the void before the quadrupole
and the profile changes qualitatively. A second peak in energy deposition appears at the entrance of the
quadrupole and eventually exceeds the first one.

\begin{figure}
  \includegraphics[width=7cm]{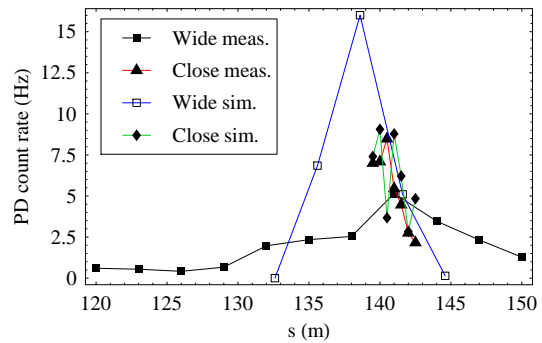}\\
  \caption{(color online) The PD signals from both configurations together with simulation results, averaged over
the fills and normalized to a typical average luminosity of $9.1\times10^{27}\mathrm{cm^{-2}s^{-1}}$. The
simulation was produced from the central BFPP orbit in the nominal lattice but with the beam pipe
misaligned by 1.4~mm at the impact location according to measurements. Due to large statistical error bars,
the simulation gives only the order of magnitude to expect in the measurement.
}\label{fig:zDist}
\end{figure}

The PDs count the number of minimum-ionizing particles (MIPs) that pass their active area of 1~cm$^2$.
In the simulation, the number of MIPs entering each PD was recorded, assuming a detection efficiency
of 30\%. Since the PDs are much more sensitive to particles entering through the side
orthogonal to the cryostat, only these particles were counted.
At the expected BFPP ion production rate of 4~kHz, and the 0.01~MIPs expected per PD per ion,
pileup from multiple interactions was not an issue. Both WPD and CPD  configurations were simulated for
various impact points and a typical result is shown together with measured data  in Fig.~\ref{fig:zDist}.

\begin{figure}
\includegraphics[width=8cm]{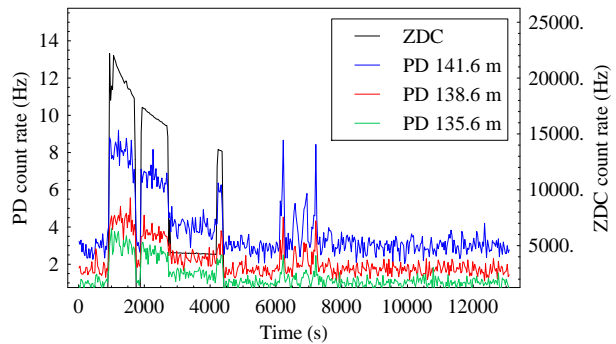}
\caption{\label{fig:Count}(color online). Count rates measured on the Zero degree calorimeter (ZDC) luminosity
monitors (black, right scale) and the three PDs with the highest signal (colors, left scale) during a
store with the WPD. The data was binned in 30~second intervals. A clear correlation between the luminosity
and the PD count rates can be seen.}
\end{figure}

\begin{figure}
  \includegraphics[width=3.9cm]{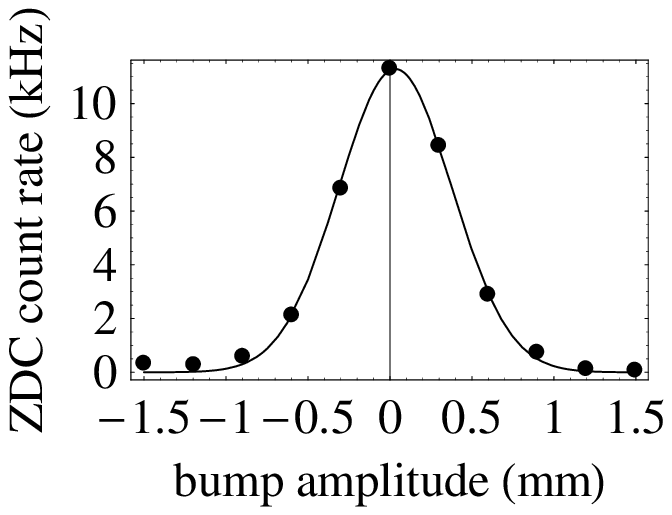}
    \includegraphics[width=4.2cm]{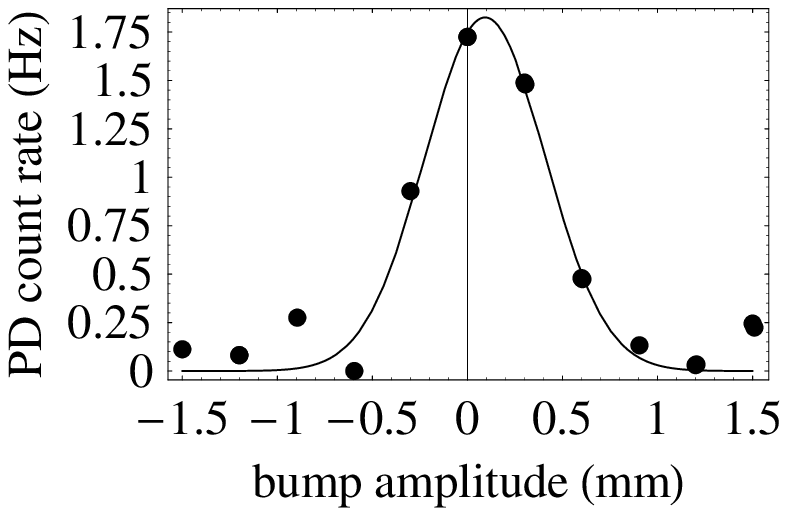}\\
\caption{The ZDC count rate (proportional to luminosity, left)
and background-subtracted PD count rate  (right)
vs. relative orbit displacement between beams in a van der Meer scan. A Gaussian fit of the
form $A \exp(\frac{-(x-x_0)^2}{\sigma_x^2})$ is shown together with the data. The two
fits had $\sigma_x = 0.49$~mm and $0.44$~mm and $\chi^2/DOF = 0.23$ and $0.20$, assuming a measurement error of 500~Hz on the ZDC signal and 0.3~Hz on the PD.}
\label{fig:vernier} \end{figure}

Data were collected from the PDs during 14~fills of RHIC with $^{63}$Cu$^{29+}$ ions.
An example of the recorded signal is shown in
Fig.~\ref{fig:Count}, together with the Zero Degree Calorimeter (ZDC)
count rate (which is directly proportional to the luminosity).
The PD signals show a good temporal correlation with luminosity and
are very localized along $s$, close to the predicted impact
position. With a 1~s binning, the correlation coefficient, $r$, between the PD signal and ZDC is 0.7 at 141.6~m and decreases gradually to 0.1 at PDs outside the predicted impact area. With a 30~s binning the highest $r$ rises to 0.98. This makes it very unlikely that some
other source than the collisions %%% than the BFPP process
could be responsible for the signals.

This becomes even  clearer from van der Meer scans
in which the beam orbits are scanned transversely
across each other and the luminosity variation is recorded as in Ref.~\cite{drees03}.
Fig.~\ref{fig:vernier} shows the ZDC rate and the signal of one
of the PDs as a function of the orbit bump amplitude.

Since  the data were taken parasitically
during normal colliding-beam operation, the PD signals were often
polluted by other losses, e.g., from collimation.
Therefore
the cleanest data sets with least interference were picked out (near
the beginning of each fill) and the averaged count rate with background
subtracted, normalized to luminosity, was calculated for each PD (see
Fig.~\ref{fig:zDist}). The background noise level was calculated for each PD as the average signal without luminosity.

In the WPD, the maximum signal was
recorded by the PD at 141.6~m but moved to 140.5~m in the CPD.
The count rates ranged from 1 to 20~Hz
depending on PD position and luminosity.  This agrees well with the simulation.
However, the maximum in the simulation came from the PD at 138.6~m, which does not
exactly agree with the measurements. This means that, in reality, the second peak in the
energy distribution outside the cryostat is actually higher than the first, while in the
simulation the first peak is the higher.  As can be seen in Fig.~\ref{fig:shower}, the peak
in the shower could also escape between the PDs in the WPD. However, if the impact point is
translated within the 2 m error bar, the simulated maximum moves further downstream and a
 fair agreement can be found.  Moreover, the PD signals themselves have error bars---apart
from the statistical error a small relative change in the counting efficiency between the PDs
could change the result.  This has not been measured so we give no numerical estimate.

One might hope to extract a value for the
cross section for BFPP between colliding copper ions at 100~GeV/nucleon. However, because of the
uncertainties in both the number
of MIPs entering a specific diode in the simulation and also the
recorded count rates, we can only conclude from the good agreement that the theoretical estimate
of the BFPP cross section has the right
order of magnitude.

No signal was detectable at 31~GeV/nucleon, consistent with
the much lower ($L\times\sigbfpp$) (see Table \ref{tab:BFPP}).

In conclusion, we made the first measurements of the localized loss of a BFPP generated secondary beam in an ion collider. These measurements were done with copper beams at RHIC.
We found the location of the maximum loss monitor signal at 140.5 m from the IP, within 2 m of its calculated location, and the measured event rates in our PD detectors within a factor of 2 of calculated ones. The deviations between calculated and measured values are consistent with estimated errors. This is a valuable test of our ability to make quantitative predictions of this effect for the LHC, where it is expected to be one of the most restrictive luminosity limits.

We thank A.J.~Baltz, R.~Gupta, K.~Hencken, J.B.~Jeanneret and T.~Roser for helpful discussions.

%\bibliography{rhic-bibl2}% Produces the bibliography via BibTeX.

\end{document}